\documentclass[12pt]{article}

\usepackage[letterpaper, margin=1in]{geometry}
\usepackage{setspace}
\usepackage{lmodern}
\usepackage[authoryear]{natbib}
\usepackage{multibib}
\newcites{SM}{Appendix References}

\usepackage{amsmath, amsfonts, amssymb, mathrsfs}
\usepackage{setspace}
\usepackage{amsmath}
\usepackage{amssymb}
\usepackage{bm}
\usepackage{multirow}
\usepackage{array}
\usepackage{booktabs}
\usepackage{mathtools}
\usepackage{array}
\usepackage{booktabs}
\usepackage{graphicx}
\usepackage{multicol, multirow}
\usepackage{pdflscape}
\usepackage{algorithm}
\usepackage{algpseudocode}
\usepackage{upgreek}
\usepackage{enumitem}
\usepackage{etoc}

\usepackage{authblk}
\usepackage{hyperref}
\hypersetup{colorlinks=true, linkcolor=blue, citecolor=blue, urlcolor=blue}

\usepackage{amsthm}

\newcommand{\EE}{\mathbb{E}}

\newcommand{\Nestedxx}{Y\big(x, M(x)\big)}
\newcommand{\Nestedxy}{Y\big(x, M(x^\ast)\big)}
\newcommand{\Nestedyy}{Y\big(x^\ast, M(x^\ast)\big)}
\newcommand{\Nestedyx}{Y\big(x^\ast, M(x)\big)}
\newcommand{\ENestedxx}{\EE\big[\Nestedxx\big]}
\newcommand{\ENestedxy}{\EE\big[\Nestedxy\big]}
\newcommand{\ENestedyy}{\EE\big[\Nestedyy\big]}
\newcommand{\ENestedyx}{\EE\big[\Nestedyx\big]}

\title{\textbf{Temperature and Respiratory Emergency Department Visits: A Mediation Analysis with Ambient Ozone Exposure}}
\author[1]{Chen Li}
\author[1]{Thomas W. Hsiao\thanks{Corresponding author: thsiao3@emory.edu}}
\author[2]{Stefanie Ebelt}
\author[1]{Rebecca H. Zhang}
\author[1,2]{Howard H. Chang}
\affil[1]{Department of Biostatistics and Bioinformatics, Emory University}
\affil[2]{Gangarosa Department of Environmental Health, Emory University}
\date{\today}
\begin{document}

\begin{singlespace}
\maketitle

\begin{abstract}
High temperatures are associated with adverse respiratory health outcomes and increases in ambient air pollution. Limited research has quantified air pollution’s mediating role in the relationship between temperature and respiratory morbidity, such as emergency department (ED) visits. In this study, we conducted a causal mediation analysis to decompose the total effect of daily temperature on respiratory ED visits in Los Angeles from 2005 to 2016. We focused on ambient ozone as a mediator because its precursors and formation are directly driven by sunlight and temperature.  We estimated natural direct, indirect, and total effects on the relative risk scale across deciles of temperature exposure compared to the median. We utilized Bayesian additive regression trees (BART) to flexibly characterize the nonlinear and non-additive relationship between temperature and ozone and quantified uncertainty via posterior prediction and the Bayesian bootstrap. We validated our modeling approach through simulation studies, which showed that BART is especially useful for estimating natural indirect effects when interactions are present. Our analysis found that ozone partially mediated the association between high temperatures and respiratory ED visits, particularly at moderately high temperatures. This study extends the existing literature by considering acute respiratory morbidity and employing a flexible modeling approach, offering new insights into the mechanisms underlying temperature-related health risks.
\end{abstract}

\vspace{1em}
\noindent \textbf{Keywords:} high temperature; ozone pollution; emergency department visits; causal mediation; Bayesian additive regression tree
\end{singlespace}

\newpage

\section{Introduction}\label{sec1}
High ambient temperature poses significant health risk globally. Studies have reported associations between short-term high temperature exposure and various adverse health outcomes, including total and cause-specific mortality \citep{RN1,RN2,RN3}, hospital encounters \citep{RN4, RN5,RN7, RN8, RN9, RN10}, and poor birth outcomes \citep{RN6}. In particular, heat exposure can exacerbate respiratory diseases, such as asthma, chronic obstructive pulmonary disease and respiratory tract infections \citep{andersonHeatrelatedEmergencyHospitalizations2013, zhuHeatExposureRespiratory2025}. Potential mechanisms include heat-induced changes in lung function \citep{RN32}, pulmonary injury due to inhaling hot air \citep{RN31}, and the use of medications that impact thermo-regulation \citep{RN33}. The number of hospitalizations and disease burden for respiratory disease attributable to extreme heat is projected to increase over the next 50 to 70 years due to climate change \citep{RN11}.

Ambient ground-level ozone is another important environmental risk factor for respiratory health. Studies have consistently shown that short-term exposure to ozone is associated with respiratory mortality and morbidity \citep{RN13,RN14}, and can further harm lung tissue, increase inflammation in the airways, and heighten the lungs' sensitivity to other irritants \citep{RN12}. Ozone is a secondary pollutant generated by sunlight-driven chemical reactions between $\text{NO}_\text{x}$ and volatile organic compounds (VOC), including methane ($\text{CH}_4$), CO and other more complex organic compounds \citep{RN15}. Ambient ozone concentrations are often highly correlated with temperature \citep{RN17}, primarily due to temperature-dependent increases in chemical reaction rates and enhanced emissions of ozone precursor compounds \citep{RN16}. As a result, ozone may act as a potential mediator on the causal pathway linking temperature to respiratory outcomes. 

Multiple studies have explored the potential mediating role of ozone under a causal mediation framework and have reported positive relative risk of indirect effects through ozone with increasing temperature \citep{RN18, RN39}. Other studies have identified positive indirect effects for ozone in the pathway between temperature and other health outcomes, including non-accidental deaths \citep{RN19}, cardiovascular diseases \citep{RN20}, and glomerular filtration rate decrease \citep{RN40}. Despite the growing body of evidence supporting the mediating role of ozone in the relationship between temperature and health, these studies all assumed a linear exposure-mediator association between temperature and ambient ozone. In addition, the outcome regression between temperature and the health outcome was also assumed to be linear and interactions between temperature and ozone were not considered. It remains unclear how well these restrictive parametric assumptions hold in practice, motivating interest in flexible nonparametric models that can achieve strong performance under minimal assumptions.  

In this study, we performed a causal mediation analysis to estimate the total, natural indirect, and natural direct effects of short-term temperature exposure on respiratory emergency department (ED) visits through ambient ozone concentration motivated by a dataset of daily respiratory ED visits in the Los Angeles metropolitan area from 2005 to 2016. We relax the strong parametric assumptions of previous analyses by using Bayesian additive regression trees (BART) to model the exposure-mediator relationship. We then compute 95\% uncertainty intervals for our mediation effect estimates through a computationally efficient Bayesian bootstrap method. We validate our method through a comprehensive simulation experiment.  

\section{Data}\label{sec2}

\subsection{Emergency Department Visits}\label{subsec1}
Daily counts of respiratory ED visits were obtained from the California Office of Statewide Health Planning and Development for the Los Angeles metropolitan area from 2005 to 2016. These records included both patients who were admitted to the hospital following an ED visit and those who were discharged directly from the ED. Respiratory ED visits were identified using primary and secondary International Classification of Diseases (ICD) diagnosis codes (ICD-9 460-519 before October 1st, 2015; ICD-10 J00-J99 afterwards). We restricted the analysis to the warm season from May to October, resulting in a total of 2,208 days.

\subsection{Meteorological Data}\label{subsec2}
Daily maximum temperature in degree Celsius and daily average absolute humidity were acquired from the High-resolution Urban Meteorology for Impacts Datasets (HUMID), a gridded dataset that provides near-surface temperature data for the contiguous U.S. at a 1km spatial resolution \citep{RN21}. This dataset explicitly accounts for urban heat islands by employing an urban canopy model using the High-Resolution Land Data Assimilation System (HRLDAS) \citep{RN23}. Bias-correction was performed using observations from various networks to improve accuracy. Using the 1km meteorology product, we first calculated a spatial average for each ZIP code overlapping the Los Angeles study area. Then, to obtain a single daily measure of exposure for the study area, a daily weighted average across ZIP codes was computed using the annual population for each ZIP code.

\subsection{Ambient Ozone Data}\label{subsec3}
Daily 8-hour maximum ozone concentrations were estimated using biased-corrected simulations of the Community Multiscale Air Quality Modeling System (CMAQ) at a 12km spatial resolution. CMAQ simulations were bias-corrected with observations from the EPA Air Quality System and land use variables \citep{RN24}. Similar to the meteorological variables, we computed daily ZIP code population weighted averages for the study area.

\section{Method}\label{sec3}

Let $X$ denote the observed exposure (temperature), $M$ the mediator (ambient ozone concentration), ${Y}$ the outcome (number of daily respiratory ED visits) and $\bm{C}$ a set of covariates (day-of-year, humidity, weekday) that adjust for confounding of both the exposure-outcome and mediator-outcome relationships. For our application, both $X$ and $M$ are continuous while $Y$ is a count variable. Let $x^\ast$ denote the reference level of $X$ and $x$ the exposed level. The unit of observation is an individual day $i$, but we suppress the date index in this section to simplify the notation. We use $f$ to denote probability densities, with subscripts specifying the variables and conditioning sets.

\subsection{Causal Assumptions}\label{model}
We use counterfactual random variables under the potential outcomes framework to define our desired causal estimands. Let $Y(x)$ denote the value $Y$ that would be observed if the exposure $X$ were set to $x$. Similarly let the nested counterfactual $\Nestedxy$ denote the outcome $Y$ that would be observed if: 1) the exposure $X$ were set to $x$, and 2) the mediator $M$ were set to the value it would attain if the exposure $X$ were set to $x^\ast$. 

For identification, we assume the standard assumptions for causal mediation analysis \citep{RN29} as follows:

\begin{enumerate}
    \item \textbf{Consistency.} If $X=x$, then $M = M(x)$, and if $M=m$, then $Y = Y(x,m)$.
    
    \item \textbf{Positivity.} For all $c$ with $f_{\bm{C}}(c) > 0$, and all $x$
    \[
    f_{X\mid \bm{C}}(x \mid c) > 0,\quad
    f_{M\mid X, \bm{C}}(m \mid x,c) > 0.
    \]
    
    \item \textbf{No unmeasured confounding for $X \rightarrow M$: $ M(x) \perp\!\!\!\perp X \mid \bm{C}.$}

    \item \textbf{No unmeasured confounding for $M \rightarrow Y$:  $Y(x,m) \perp\!\!\!\perp M \mid \bm{C}.$}

    \item \textbf{No unmeasured confounding for $X \rightarrow Y$:  $Y(x,m) \perp\!\!\!\perp X \mid \bm{C}.$}
    
    \item \textbf{Cross-world independence: $Y(x,m) \perp\!\!\!\perp M(x^\ast) \mid \bm{C}.$}

\end{enumerate}

\subsection{Causal Estimands and Identification}\label{model}

We now define our five causal estimands of interest. We used the risk ratio scale rather than the risk difference to quantify our causal mediation effects. The pure natural direct effect (PNDE) and the total natural direct effect (TNDE) are defined as
\[ PNDE = \frac{\ENestedxy}{\ENestedyy}, \qquad \qquad TNDE = \frac{\ENestedxx}{\ENestedyx} \;.\]
PNDE describes how the value of $Y$ changes when exposure $X$ increases from the reference level $x^\ast$ to $x$, while keeping the mediator $M$ at the same value that it would attain when $X = x^\ast$. In contrast, TNDE describes effects of changing $X$ from $x^\ast$ to $x$, while keeping $M$ at the same value that it would attain under $X = x$. PNDE and TNDE are identical when there is no exposure-mediator interaction. Similarly, the pure natural indirect effect (PNIE) and total natural indirect effect (PNIE) can be defined as:
\[ PNIE = \frac{\ENestedyx}{\ENestedyy}, \qquad \qquad TNIE = \frac{\ENestedxx}{\ENestedxy} \;. \]
Finally, the total effect (TE) is defined as: 
\[ TE = \frac{\ENestedxx}{\ENestedyy} \;, \]
where $TE = TNDE \times PNIE$ or alternatively $TE= PNDE \times TNIE$.

All the defined causal estimands are simple ratios of the mean nested counterfactual with different combinations of exposure and mediator contrasts. By our causal assumptions, identification follows by the nonparametric mediation $g$-formula \citep{RN25}:
\begin{equation}
\ENestedxy = \int\int \EE(Y \mid X = x, M = m, \bm{C}=\bm{c}) \, f_{M\mid X=x^\ast, \bm{C}}(m \mid  \bm{c}) \, f(\bm{c}) \, dm\,d\bm{c}.
\label{eq1}
\end{equation}

\subsection{Estimation and Inference}\label{model}

The identification formula \eqref{eq1} suggests that inference can proceed by estimation of two nuisance functions: 1) $\EE(Y \mid X=x, M=m, \bm{C}=\bm{c})$, the \textit{outcome regression}, and 2) $f_{M|X,\bm{C}}(m\mid \bm{c})$, the \textit{mediator density}.

\subsubsection{Outcome regression}

We assumed the outcome regression follows the established quasi-Poisson log-linear model for time-series analysis of temperature and hospital encounter counts \citep{RN34}, given by 
\begin{equation}
\log \big\{E( Y | X=x, M=m, \bm{C}=\bm{c})\big\} = \theta_0 + f(x,\bm{\theta_1}) + \theta_2 m + \sum_{h=1}^{3} \theta_{3,h} I_h(x)\times m + \bm{\theta_4^\top c}\,.
\label{eq2}
\end{equation}
 The temperature effect $f(x,\bm{\theta_1})$ was modeled non-linearly using natural cubic splines with 6 degrees of freedom and $\bm{\theta_1}$ is the vector of basis coefficients. The ozone main effect $\theta_2$ was assumed to be linear since there is limited evidence of non-linearity from prior epidemiologic studies. To capture the interaction between temperature and ozone concentration, we categorized temperature into four quartile categories. Here $I_h(x)$ denotes an indicator for the $(h+1)^{\text{th}}$ quartile of the overall distribution of temperature $X$ (evaluates to 1 if $x$ falls in that quartile and 0 otherwise). Other confounders in the health model included natural cubic splines for day-of-year with 6 degrees of freedom and their interactions with indicators for years, natural cubic splines for specific humidity with 6 degrees of freedom, indicators for weekday, and an indicator for federal holidays. 

\subsubsection{Mediator density}
 In standard causal inference, one would nonparametrically estimate the mediator density conditional on $X$ and $\bm{C}$. However, because $M$ is continuous and $\bm{C}$ is multidimensional, direct estimation of this conditional density can be unstable due to the curse of dimensionality. At the same time, we would prefer to avoid imposing strong parametric assumptions on the relationship among $M, X,$ and $\bm{C}$. One approach that complements the quasi-Poisson log-linear outcome regression shifts the nuisance estimation burden from the conditional density to the conditional expectation $\EE[M\mid X=x, \bm{C}=\bm{c}]$, while still permitting estimation of \eqref{eq1}. To see why, we substitute \eqref{eq2} into \eqref{eq1} and assume that the conditional distribution of $M$ given $X$ and $\bm{C}$ is Gaussian with constant variance $\sigma^2$. Under this assumption, it follows that
 \begin{equation}
   \begin{split}
    &\ENestedxy = \int\int \EE(Y \mid X = x, M = m, \bm{C}=\bm{c}) \, f_{M\mid X=x^\ast, \bm{C}}(m \mid  \bm{c}) \, f(\bm{c}) \, dmd\bm{c} \\
     &= \exp\{\theta_0 + f(x, \boldsymbol{\theta_1})\}\int \exp\{\boldsymbol{\theta_{4}^\top c}\} \int \exp{(\theta_2m+\theta_{3x}m)}\times f_{M\mid X=x^\ast, \bm{C}}(m \mid  \bm{c}) \ dm \, f(\bm{c}) \, d\bm{c} \\
     &= \exp\{\theta_0 + f(x, \boldsymbol{\theta_1})\}\int \exp\{\boldsymbol{\theta_{4}^\top c}\} \EE\big[ \exp\{\theta_2m+\theta_{3x}m)\} \mid X = x^\ast, \bm{C}=\bm{c}\big] \, f(\bm{c}) \, d\bm{c} \\
     &= \exp\left\{\theta_0 + f(x, \boldsymbol{\theta_1}) + \frac{1}{2} (\theta_2+\theta_{3x})^2\sigma^2\right\} \EE_{\bm{C}}\bigg[ \exp\big\{\boldsymbol{\theta_{4}^\top C} + (\theta_2 + \theta_{3x}) \EE[M \mid X=x^\ast, \bm{C}]\big\}  \bigg].
    \end{split}
    \label{eq:identification}
\end{equation}

This new identification formula suggests that under our assumptions on the outcome regression, we can estimate $\EE[M\mid X, C]$ and $\sigma^2$ in place of the mediator density. We propose two methods to do this. Our main method uses Bayesian additive regression trees \citep{RN27}, or BART,  to capture nonlinear and non-additive relationships between the mediator $M$ and exposure $X$. BART is a Bayesian non-parametric approach that assumes
 \begin{equation}
     M= \sum_{g=1}^G \mathcal{T}_g (X, \bm{C})+\epsilon, 
     \qquad \epsilon \sim N(0,\sigma^2),
     \label{eq:bart}
 \end{equation}
 using a sum of $G$ decision trees. Each $\mathcal{T}_g$ is composed of a tree structure that encodes binary splits of covariates and a set of terminal leaf node. Priors are designed to favor shallow trees and shrinkage across leaf nodes. We assumed the residual errors for $M$ to be Gaussian and included all confounders $\bm{C}$ used in the health model as covariates in BART. As a parametric alternative for comparison, we also fitted a traditional additive spline model: 
 \begin{equation}
M = \beta_0 + g(X,\bm{\beta_1}) + \bm{\beta_2}^\top\bm{C}+\epsilon, 
     \qquad \epsilon \sim N(0,\sigma^2).
     \label{eq:linear}
 \end{equation}
 The temperature effect $g(x,\bm{\beta_1})$ was modeled nonlinearly using natural cubic splines with 6 degrees of freedom and $\bm{\beta_1}$ is the vector of basis coefficients.
 
Both the BART and additive spline methods allow us to estimate both $\EE[M\mid X, C]$ and $\sigma^2$ to input into \eqref{eq:identification}. Formulas for the five causal estimands under the identification in \eqref{eq:identification}, and the BART and additive spline methods in \eqref{eq:bart} and \eqref{eq:linear} are given in Appendix~\ref{sec:appendix_estimation}.

\subsubsection{Estimation and Uncertainty Quantification}

\begin{algorithm}
\caption{BART-Based Estimation of $\mathbb{E}\{Y(x,M(x^\ast))\}$ and Natural Effects}
\label{alg:bart_mediation}

\centering
\scalebox{0.85}{%
\begin{minipage}{1.22\textwidth}
\begin{algorithmic}[1]
\Require Observed data $\{Y_t, X_t, M_t, \mathbf{C}_t\}_{t=1}^T$, exposure levels $x$ and $x^\ast$, and number of draws $K$
\Ensure Posterior draws and 95\% CIs for $\mathbb{E}[Y(x,M(x^\ast))]$ and derived natural effects

\State \textbf{Outcome model and identification.}
  \State Fit the quasi-Poisson log–linear outcome model in \eqref{eq2} and obtain $\hat{\boldsymbol{\theta}}$ and $\hat{\Sigma}=\widehat{\mathrm{Var}}(\hat{\boldsymbol{\theta}})$. Under the Gaussian working model $M \mid X,\mathbf{C}\sim \mathcal{N}( \mathbb{E}[M\mid X,\mathbf{C}],\sigma^2)$:
  \begin{equation*}
  \begin{split}
    \mathbb{E}[Y(x,M(x^\ast))]
    &= \exp\!\left\{\theta_0 + f(x,\boldsymbol{\theta}_1) 
          + \tfrac{1}{2}(\theta_2+\theta_{3x})^2\sigma^2 \right\} \\
    &\quad \times \mathbb{E}_{\mathbf{C}}\!\left[
        \exp\big\{ \boldsymbol{\theta}_4^\top \mathbf{C}
                   + (\theta_2+\theta_{3x})\,\mathbb{E}[M \mid X=x^\ast,\mathbf{C}] \big\}
      \right].
  \end{split}
  \end{equation*}

\State \textbf{Step 1: Draw coefficient samples for the outcome model.}
  \State For $k = 1,\dots,K$ (e.g., $K=5{,}000$), draw
  \[
    \boldsymbol{\theta}^{(k)} \sim \mathcal{MVN}(\hat{\boldsymbol{\theta}},\hat\Sigma).
  \]

\State \textbf{Step 2: Fit BART mediator model and obtain $\mathbb{E}[M\mid X=x^\ast,\mathbf{C}]$ and $\sigma^2$.}
  \State Fit a BART model for $\mathbb{E}[M\mid X,\mathbf{C}]$ and $\sigma^2$ using the observed data, 200 trees and a burn-in of 5{,}000 iterations. For each posterior draw $k$ and each day $t=1,\dots,T$:
  \[
    \hat m_{|x^\ast}^{(k,t)} \sim \mathcal{N}(\sum_{g=1}^G \widehat{\mathcal{T}}_g (X = x^*, \bm{C}_t), \hat{\sigma}^2).
  \]

\State \textbf{Step 3: Compute conditional counterfactual means for each $(k,t)$.}
  \For{$k = 1,\dots,K$}
    \For{$t = 1,\dots,T$}
      \State Using $\boldsymbol{\theta}^{(k)}$ and $\hat m_{|x^\ast}^{(k,t)}$, compute
      \begin{align*}
        \hat{\mathcal{F}}(x,x^\ast \mid \mathbf{C}_t)^{(k)}
        &= \exp\!\Big\{
              \hat\theta_0^{(k)}
              + f\big(x;\hat{\boldsymbol{\theta}}_1^{(k)}\big)
              + (\hat{\boldsymbol{\theta}}_4^{(k)})^\top \mathbf{C}_t \\
        &\qquad\qquad\quad
              + (\hat\theta_2^{(k)}+\hat\theta_{3x}^{(k)})\,\hat m_{|x^\ast}^{(k,t)}
              + \tfrac{1}{2}\big(\hat\theta_2^{(k)}+\hat\theta_{3x}^{(k)}\big)^2
                \hat\sigma^2
            \Big\}.
      \end{align*}
    \EndFor
    
\State \textbf{Step 4: Bayesian Bootstrap over $\mathbf{C}$.}
    \State Draw weights as $\mathbf{w}^{(k)} = (w_1^{(k)},\dots,w_T^{(k)})^\top 
      \sim \mathrm{Dirichlet}(1,\dots,1)$.
    \State Form the Bayesian Bootstrap–weighted average:
    \[
      \hat{\mathcal{F}}(x,x^\ast)_{\mathrm{avg}}^{(k)}
      = \sum_{t=1}^T w_t^{(k)} \,
        \hat{\mathcal{F}}(x,x^\ast \mid \mathbf{C}_t)^{(k)}.
    \]
    \State Repeat the same steps to obtain $\hat{\mathcal{F}}(x,x)_{\mathrm{avg}}^{(k)}, \hat{\mathcal{F}}(x^\ast,x^\ast)_{\mathrm{avg}}^{(k)}$, and $\hat{\mathcal{F}}(x^\ast,x)_{\mathrm{avg}}^{(k)}$.
  \EndFor

\State \textbf{Step 5: Construct posterior draws for natural effects.}
  \State For example, the pure natural direct effect (PNDE) on the multiplicative scale is
  \[
    \mathrm{PNDE}^{(k)}
    = \frac{\hat{\mathcal{F}}(x,x^\ast)_{\mathrm{avg}}^{(k)}}
           {\hat{\mathcal{F}}(x^\ast,x^\ast)_{\mathrm{avg}}^{(k)}},
    \qquad k=1,\dots,K.
  \]

\State \textbf{Step 6: Point estimates and 95\% intervals.}
  \State  Posterior mean and empirical 0.025 and 0.975 quantiles of  $\{\mathrm{PNDE}^{(k)}\}_{k=1}^K$ as the 95\% interval bounds.
\end{algorithmic}
\end{minipage}
}
\end{algorithm}

Uncertainty for the estimators based on \eqref{eq:identification} was assessed by Monte Carlo simulations (Algorithm~\ref{alg:bart_mediation}). We first generated $K = 5,000$ sets of coefficients $\bm{\theta}^{(k)}$ from the asymptotic distribution of the  outcome regression \eqref{eq2}. From the BART model, we specified 200 trees with a burn-in period of 5,000 iterations and generated $K$ samples of $M^{(k)}$ from the posterior prediction distributions. For each $k^\text{th}$ iteration, we plugged $\bm{\theta}^{(k)}$ and the prediction of $M$ into \eqref{eq:identification} for each observation day to obtain the expectation of the counterfactual variables. Second, we accounted for the variability of the confounders $\bm{C}$ by performing a Bayesian Bootstrap. Instead of fitting BART on multiple bootstrap replications generated by a parametric bootstrap, we assigned different weights to each observation in different iterations. Specifically, for each iteration $k$, we generated weights for every observation day from a Dirichlet distribution with parameters $w_1 = \cdots = w_T = 1$, followed by averaging the weighted estimated expectation of counterfactual variables over all the observation days. Finally, we define the point estimate of the causal mediation effect as the mean of the nested counterfactual variables across iterations and the 95\% confidence interval bounds are given by the 2.5th and the 97.5th quantiles. For the conventional additive spline mediator regression, we obtained a closed-form estimate of $\EE[M \mid X=x^\ast, \bm{C}]$ and a parametric bootstrap was used to obtain the confidence interval. Additional details are given in Appendix~\ref{sec:appendix_estimation}.

\section{Simulation Study}\label{sec4}

We conducted a simulation study designed to resemble the observed dataset to evaluate estimation performance of the proposed mediation analysis. First, we followed the general structure of the observed temperature and other confounders in the Los Angeles application to develop two true mediation models: an additive model with no interactions between the exposure and the covariates, and the same model but with an interaction effect between temperature and humidity. The interaction model follows
\begin{equation*}
\begin{aligned}
\mathbb{E}(M \mid X=x, \bm{C}=\bm{c})
&=
\mu_{\mathrm{add}}(x,\bm{c})
+\left[
0.65\,I(x>q_{0.65})
+0.35\,I(x>q_{0.80})
\right]G(h),\\
G(h)
&=
\begin{cases}
+0.5, & h>\operatorname{median}(H),\\
-0.5, & h\leq\operatorname{median}(H).
\end{cases}
\end{aligned}
\end{equation*}
where $\mu_{\mathrm{add}}(x,c)$ is the conditional mean for the additive model, $q_a$ represents the $a$th quantile for the temperature variable, $I(\cdot)$ is an indicator function, and $h$ is the humidity covariate, which is contained in the covariate vector $\bm{C}$. Details on the additive model and full simulation can be found in Appendix~\ref{sec:appendix_sim}. 

Second, we simulated daily respiratory ED visits from a negative binomial distribution where coefficients in the health models were estimated from the real data. Finally, we estimated PNDE, TNIE, PNIE, TNDE, and TE of temperature effect on ED visits by using either the BART \eqref{eq:bart} or spline estimator \eqref{eq:linear} to estimate the temperature-ozone relationship $\mathbb{E}(M\mid X,\bm{C})$. We calculated percent relative bias (\%RB), root mean square error (RMSE) and 95\% interval coverage over 500 simulations. The pointwise 95\% uncertainty intervals were computed according to the procedure in Algorithm~\ref{alg:bart_mediation} with $K=5{,}000$ draws.

\begin{table}[!b]
\caption{Relative bias (RB), root mean squared error (RMSE), and coverage (\%) for PNDE, TNIE, TE, PNIE, and TNDE across simulation scenarios. Coverages that fell below 90\% are bolded (expected is 95\%). \label{tab:pi_rmse_new}}
\renewcommand{\arraystretch}{2}
\resizebox{\textwidth}{!}{%
\begin{tabular}{c l l ccc ccc ccc ccc ccc}
\toprule
\multirow{2}{*}{\textbf{Exposure}} &
\multirow{2}{*}{\textbf{$f_{M \mid X, \bm{C}}$}} &
\multirow{2}{*}{\textbf{Model}} &
\multicolumn{3}{c}{\textbf{PNDE}} &
\multicolumn{3}{c}{\textbf{TNIE}} &
\multicolumn{3}{c}{\textbf{TE}} &
\multicolumn{3}{c}{\textbf{PNIE}} &
\multicolumn{3}{c}{\textbf{TNDE}} \\
\cmidrule(lr){4-6}
\cmidrule(lr){7-9}
\cmidrule(lr){10-12}
\cmidrule(lr){13-15}
\cmidrule(lr){16-18}
& & &
\textbf{\%RB} & \textbf{RMSE} & \textbf{Coverage} &
\textbf{\%RB} & \textbf{RMSE} & \textbf{Coverage} &
\textbf{\%RB} & \textbf{RMSE} & \textbf{Coverage} &
\textbf{\%RB} & \textbf{RMSE} & \textbf{Coverage} &
\textbf{\%RB} & \textbf{RMSE} & \textbf{Coverage} \\
\midrule

\multirow{4}{*}{0.75 vs 0.50}
& Additive
& Spline
&  0.006 & 0.004 & 94.8
& -0.003 & 0.001 & 93.6
&  0.003 & 0.004 & 94.8
& -0.005 & 0.001 & 94.0
&  0.008 & 0.004 & 95.4 \\

& Additive
& BART
&  0.006 & 0.004 & 94.8
& -0.007 & 0.001 & 94.6
& -0.001 & 0.004 & 94.6
& -0.009 & 0.001 & 94.8
&  0.008 & 0.004 & 95.2 \\

& Interaction
& Spline
&  0.025 & 0.004 & 94.2
&  0.086 & 0.001 & \textbf{84.4}
&  0.111 & 0.004 & 94.6
&  0.088 & 0.001 & \textbf{87.6}
&  0.024 & 0.004 & 94.4 \\

& Interaction
& BART
&  0.026 & 0.004 & 94.2
& -0.005 & 0.001 & 95.6
&  0.021 & 0.004 & 95.0
&  0.000 & 0.001 & 97.0
&  0.021 & 0.004 & 94.6 \\

\midrule

\multirow{4}{*}{0.85 vs 0.50}
& Additive
& Spline
& -0.006 & 0.003 & 94.0
& -0.009 & 0.001 & 96.8
& -0.015 & 0.003 & 94.6
& -0.004 & 0.001 & 94.0
& -0.011 & 0.004 & 94.8 \\

& Additive
& BART
& -0.008 & 0.003 & 94.2
& -0.021 & 0.001 & 95.8
& -0.029 & 0.003 & 94.0
& -0.034 & 0.001 & 94.0
&  0.005 & 0.004 & 95.0 \\

& Interaction
& Spline
&  0.044 & 0.003 & 96.2
&  0.055 & 0.002 & 91.6
&  0.099 & 0.004 & 93.6
&  0.153 & 0.002 & \textbf{83.8}
& -0.054 & 0.004 & 94.4 \\

& Interaction
& BART
&  0.028 & 0.003 & 95.8
& -0.002 & 0.001 & 95.2
&  0.026 & 0.003 & 95.6
&  0.009 & 0.001 & 97.6
&  0.018 & 0.004 & 95.2 \\

\midrule

\multirow{4}{*}{0.95 vs 0.50}
& Additive
& Spline
&  0.005 & 0.004 & 95.2
& -0.015 & 0.002 & 96.4
& -0.011 & 0.004 & 95.6
& -0.007 & 0.002 & 94.6
& -0.003 & 0.005 & 96.0 \\

& Additive
& BART
&  0.003 & 0.004 & 94.8
& -0.016 & 0.002 & 96.6
& -0.014 & 0.004 & 95.6
& -0.012 & 0.002 & 94.8
& -0.002 & 0.005 & 96.0 \\

& Interaction
& Spline
&  0.048 & 0.004 & 97.0
&  0.058 & 0.002 & 93.0
&  0.104 & 0.004 & 94.8
&  0.169 & 0.003 & \textbf{89.2}
& -0.063 & 0.005 & 94.6 \\

& Interaction
& BART
&  0.032 & 0.004 & 97.2
&  0.005 & 0.002 & 95.4
&  0.037 & 0.004 & 96.4
&  0.037 & 0.002 & 94.6
&  0.001 & 0.005 & 95.6 \\

\bottomrule
\end{tabular}
}
\end{table}

Table 1 presents the effects of temperature at three quantile levels (0.75, 0.85, 0.95) compared to the median temperature. Under the additive scenario, fitting the data with BART resulted in comparable or improved error to that of the spline estimator and coverage for each method was reasonably close to the nominal 95\% level. For the scenario with interaction effect, estimation with BART yielded better \%RBs and RMSEs. In addition, coverage of BART stayed at the nominal $95\%$ level but coverage for the spline method for some estimands fell to as low as $83.8\%$. 

The worse performance of the spline method was especially pronounced when the causal estimand was an indirect effect (PNIE or TNIE) for the $0.75$ vs. $0.50$ and $0.85$ vs $0.50$ ratios. The \%RB for the spline estimate of the 0.75 vs. 0.50 TNIE was $0.086$ with a coverage of $84.4\%$ whereas BART had \%RB of only $-0.005$ with $95.6\%$ coverage. For PNIE of $0.85$ vs. $0.50$, spline had \%RB of $0.153$ and coverage of $83.8\%$, and BART had \%RB of $0.009$ and $97.6\%$ coverage. The improved estimation properties of BART for the indirect effects (and by extension the TE) can be traced back to the definition of the five causal estimands and the decomposition of the marginal causal mean in \eqref{eq:identification}. For direct effects like PNDE and TNDE, the mediator distribution is the same between the numerator and denominator so that even if the exposure-mediator model is poorly fit, the errors will partly cancel. In contrast, the indirect effects are defined by different mediator distributions in the numerator and denominator, so that errors in fitting the exposure-mediator relationship propagate much more heavily. However, the estimation of the misspecified spline model does improve for the 0.95 vs. 0.50 indirect effects. This is likely just a feature of our simulation, because the thresholds for the interaction effects ($q_{0.65}$ and $q_{0.80}$) are closer to the 0.75 and 0.85 quantiles than the 0.95 quantile, so that estimation error is much more sensitive within neighborhoods of these exposure levels. 

Overall, the BART approach outperformed or matched the spline model across all evaluation metrics for causal effect estimation, highlighting its flexibility and robustness relative to traditional additive spline regression for estimating mediation effects for nonlinear and non-additive data.

\section{Application: Los Angeles Respiratory Emergency Department Visit Analysis}\label{sec5}

The total number of respiratory ED visits in Los Angeles during 2005-2016 was 345,922. The daily respiratory ED visit count ranged from 965 to 3,124, with a median of 1,548 visits per day. The mean of daily maximum temperature was 29.01 Celsius, with a standard deviation (SD) of 4.12. The mean of daily specific humidity was 9.33 g/kg (SD = 1.50 g/kg). The mean of daily ambient ozone concentration was 0.048 ppm (SD= 0.009 ppm). We define the reference level $x^\ast$ as the median (50th percentile) temperature.  

\vspace{-8pt}
\begin{figure}[H] 
    \centering
    \includegraphics[width=0.8\textwidth]{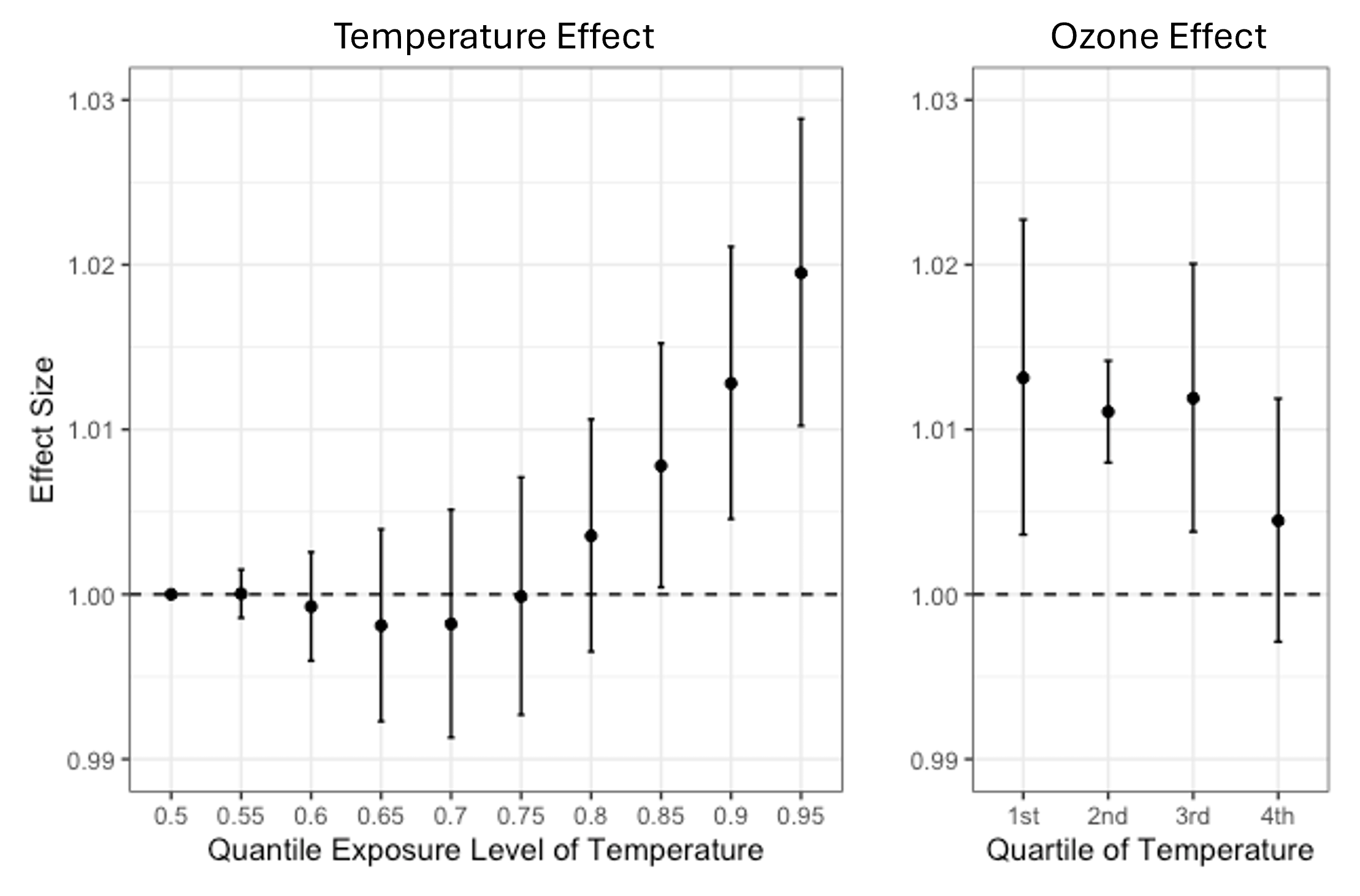}
    \caption{Relative risks of respiratory ED visits comparing daily maximum temperature at different quantile levels with the reference temperature set at the 0.50 quantile (Left Panel) and for interquartile range increases in ambient ozone concentration at different quartiles of temperature (Right Panel). Intervals indicate 95\% confidence intervals.}
    \label{fig:healthEffects}
\end{figure}
\vspace{-8pt}

Figure 1 shows the relative risks associated with temperature and ozone from the health models. We found a non-linear effect of temperature with stronger associations observed when temperature exceeds the 0.85 quantile compared to the median. We also found significant associations between ozone and respiratory ED visits when temperature was at the 1st, 2nd, and 3rd quartiles.
\begin{table}[!t]
\caption{Causal mediation effects and 95\% confidence intervals (CI) for temperature deciles compared to the median temperature (PNDE: pure natural direct effect; TNDE: total natural direct effect; PNIE: pure natural indirect effect; TNIE: total natural indirect effect; TE: total effect).\label{tab:mediation}}
\renewcommand{\arraystretch}{1.4} 
\resizebox{\textwidth}{!}{%
\begin{tabular}{lccccc}
\toprule
\textbf{Exposure Level} & \textbf{PNDE} & \textbf{TNDE} & \textbf{PNIE} & \textbf{TNIE} & \textbf{TE} \\
\midrule
0.55 & 1.0000 (0.9986, 1.0015) & 1.0000 (0.9986, 1.0015) & 1.0005 (1.0000, 1.0022) & 1.0005 (1.0000, 1.0022) & 1.0005 (0.9987, 1.0025) \\
0.60 & 0.9993 (0.9959, 1.0026) & 0.9993 (0.9959, 1.0026) & 1.0006 (1.0000, 1.0023) & 1.0006 (1.0000, 1.0023) & 0.9998 (0.9963, 1.0033) \\
0.65 & 0.9981 (0.9922, 1.0040) & 0.9981 (0.9922, 1.0040) & 1.0016 (1.0004, 1.0032) & 1.0016 (1.0004, 1.0032) & 0.9997 (0.9939, 1.0055) \\
0.70 & 0.9982 (0.9912, 1.0052) & 0.9982 (0.9912, 1.0052) & 1.0019 (1.0005, 1.0038) & 1.0019 (1.0005, 1.0038) & 1.0001 (0.9931, 1.0070) \\
0.75 & 0.9971 (0.9892, 1.0051) & 0.9959 (0.9872, 1.0046) & 1.0021 (1.0005, 1.0041) & 1.0008 (0.9995, 1.0023) & 0.9979 (0.9899, 1.0061) \\
0.80 & 1.0008 (0.9937, 1.0079) & 0.9991 (0.9912, 1.0071) & 1.0027 (1.0008, 1.0051) & 1.0010 (0.9994, 1.0029) & 1.0018 (0.9946, 1.0091) \\
0.85 & 1.0051 (0.9983, 1.0118) & 1.0032 (0.9958, 1.0106) & 1.0029 (1.0008, 1.0055) & 1.0011 (0.9993, 1.0032) & 1.0062 (0.9995, 1.0128) \\
0.90 & 1.0100 (1.0030, 1.0172) & 1.0075 (0.9999, 1.0151) & 1.0041 (1.0012, 1.0069) & 1.0016 (0.9990, 1.0042) & 1.0116 (1.0049, 1.0183) \\
0.95 & 1.0167 (1.0089, 1.0245) & 1.0129 (1.0045, 1.0213) & 1.0054 (1.0029, 1.0081) & 1.0021 (0.9994, 1.0048) & 1.0191 (1.0119, 1.0261) \\
\bottomrule
\end{tabular}
}
\end{table}
The results from our causal mediation analysis are shown in Table 2. We focus on the results for the 0.95 quantile exposure level. For the pure natural direct effect, the respiratory ED visit number increases by 1.67\% (95\% CI: 0.89\%, 2.45\%) when the temperature increases from the reference median level to the exposure level, assuming that ambient ozone concentration is fixed at the level it would be when the temperature is at reference level. For the total natural direct effect, the respiratory ED visit number increases by 1.29\% (95\% CI: 0.45\%, 2.14\%) when the temperature increases from the reference to the exposure level, assuming that ambient ozone concentration is fixed at the level it would be when the temperature is at the exposed level. For the pure natural indirect effect, the temperature remains fixed at the reference level, but the ambient ozone concentration changes from its value when the temperature is at the reference level (0.50 quantile) to the value when the temperature is at the exposed level (0.95 quantile). This change in ambient ozone exposure results in a 0.54\% (95\% CI: 0.29\%, 0.81\%) increase in respiratory ED visit counts. For the total natural indirect effect, the respiratory ED visit number is estimated to increase by 0.21\% (95\% CI: -0.06\%, 0.48\%) when the ambient ozone concentration changes from the value when the temperature is at the reference level (0.50 quantile) to the value when the temperature is at the exposed level (0.95 quantile), assuming that the temperature is fixed at the exposed level. For the total effect, the respiratory ED visit number increases by 1.91\% (95\% CI: 1.19\%, 2.61\%) when temperature increases from the reference to the exposed level. 

\begin{figure}[!t] 
    \centering
    \includegraphics[width=0.8\textwidth]{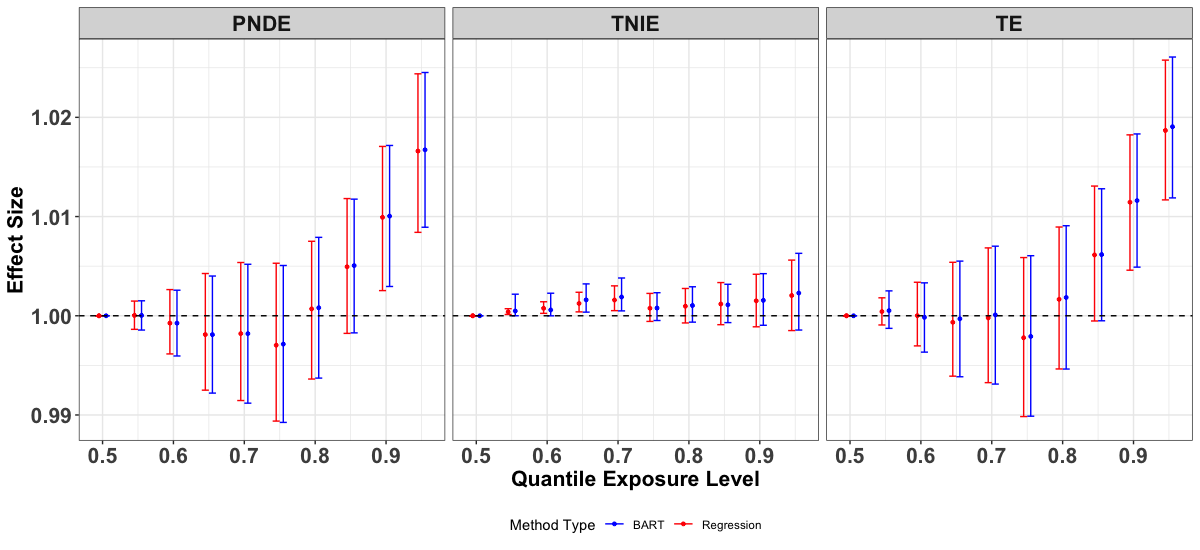}
    \caption{Comparison of causal mediation effects and 95\% confidence interval (CI) for temperature deciles compared to the median temperature between the BART approach and the regression-based approach (PNDE: pure natural direct effect; PNIE: pure natural indirect effect; TE: total effect).}
    \label{fig:mediation_plot_compare}
\end{figure}

Results for the PNDE, TNIE, and TE are shown in Figure~\ref{fig:mediation_plot_compare} to illustrate how mediation effects vary nonlinearly across exposure levels. As temperature increased from the reference median level to higher levels, the TE became positive at elevated temperatures and increased in magnitude beyond approximately the 0.75 quantile. The PNDE followed a nearly identical pattern. In contrast, the TNIE was small across most exposure levels and was distinguishable from the null primarily at moderate temperature quantiles (approximately 0.65–0.75), although its point estimates were positive beginning around the 0.60 quantile.

These results indicate a shift in the way temperature affects ED visits across the exposure range. At moderate temperatures, temperature affects ED visits partly through changes in ozone concentration. At higher temperatures, however, the increase in the TE is driven mainly by the direct effect of temperature, with minimal contribution from ozone mediation.

Additionally, we compared our results from the BART approach with those from the regression-based approach with parametric bootstrap confidence intervals (Figure~\ref{fig:mediation_plot_compare}). The point estimates and estimated confidence intervals of direct effects from these two methods were very similar. Although minor discrepancies were observed in the estimation of indirect effects between the two methods, the results were overall consistent. Compared to our BART method, the regression-based approach resulted in narrower confidence intervals for the indirect effect when the exposure level was low. 

\section{Discussion}\label{sec6}
Using our proposed method, we analyzed 12 years of respiratory-related ED visit records in Los Angeles during the warm season. We provide identification conditions and estimation procedures for natural direct effects of temperature on ED visits and natural indirect effects through ambient ozone concentrations. Our simulation study highlights the advantages in point estimation and uncertainty quantification using BART to estimate the exposure-mediator relationship when the data generating process is complex (nonlinear or non-additive), particularly for natural indirect effects (PNIE, TNIE). Our analysis provides additional epidemiological evidence that ambient ozone partially mediates the association between high temperature and increased respiratory-related ED visits. For example, at the 0.95th quantile exposure level for temperature, we reported a total natural indirect effect (TNIE) increase of 0.21\% (95\% CI: -0.06\%, 0.48\%) for ozone and a 1.91\% (95\% CI: 1.19\%, 2.61\%) increase in the total effect (TE). These results help elucidate the potential impact of ambient ozone on the effect of heat on health outcomes.

Though there has been no study conducted specifically on the causal mediation effects of ozone on the association between temperature and respiratory morbidity as measured by ED visits, our results are consistent with a previous study on respiratory mortality. A study in French urban areas estimated the pooled NIE relative risk for ozone being 1.04 (95\% CI: 1.00, 1.07) based on a binary exposure variable indicating whether or not the day fell within a heat wave (Alari et al., 2023). A study in London also identified a positive relative risk for ozone of 1.009 (95 \% CI: 1.000, 1.022) based on a binary exposure of occurrence of heatwave events \citep{RN39}. Moreover, our results consolidate the finding that ozone positively mediates the relationship between high temperature and various health outcomes. In the France study, positive mediating effects of ozone were identified on the relationship between high temperature and non-accidental and cardiovascular mortality as 1.03 (95\% CI: 1.02, 1.05) and 1.03 (95\% CI: 1.01, 1.04). Another study conducted in South Korea found that indirect effects through increased ozone were 1.0002 (95\%: 0.9999, 1.0004) and 1.0003 (95\% CI: 1.0002, 1.0005) on days with higher than or lower than minimum mortality temperature, respectively, based on a moving average of daily mean temperature as the exposure variable and non-accidental death as the outcome \citep{RN19}. Additionally, a study in China also suggests that the indirect effect of temperature on ischemic heart disease mortality through ozone was 1.18 \citep{RN20}.

In conclusion, our results show that the effect of high temperatures on respiratory-related ED visits can be explained partially through ozone, meaning that high temperatures not only affect human respiratory health by exposing the population to heat but also by generating ozone, particularly on days with moderately high temperature. This study provides insight into the potential role and mechanism driving the effect of high temperatures on population health.

Our study has several strengths. First, to the best of our knowledge, our analysis is the first ozone mediation analysis to study temperature's effect on respiratory-related ED visits as opposed to mortality. Compared to mortality, ED visits are more immediate and widespread measure of acute morbidity, and provide an improved understanding of the burden on healthcare systems. Our analysis is also the first to account for a nonlinear temperature effect in the outcome regression.

Second, we developed a more flexible and robust method to estimate causal mediation effects by integrating BART into the mediator regression. Previous mediation analyses relied on parametric linear \citep{RN18, RN19, RN39} or Poisson regressions \citep{RN20} with additive effects to model the association between temperature and ozone. BART combines the strengths of machine learning and Bayesian inference, allowing it to capture nonlinear relationships and complex interactions while providing uncertainty quantification \citep{RN26}. In the literature, \cite{lineroMediationAnalysisUsing2025} and \cite{tingEstimatingHeterogeneousCausal2025} have used BART for binary exposures to directly model and regularize heterogeneous direct and indirect effects. In contrast, our study focuses on direct and indirect effects for a continuous exposure in temperature where BART is mainly used as a nonparametric estimator of the exposure-mediator relationship only, preserving interpretability in the scientifically informed outcome regression. Our nonparametric sum-of-trees model flexibly estimates the mean structure, enabling the temperature-ozone association to differ by humidity, seasonality, and time period \citep{RN27}. 

Third, we included interaction terms between temperature and ambient ozone in the outcome regression. Existing studies suggest interactions between temperature and ozone on health outcomes \citep{RN28} and we identified significant interaction effects in our application as well. Omitting the interaction in the model can result in incorrect specification of the health model, resulting in bias in the assessment of causal mediation effects \citep{RN29}. Reporting both the PNIE and TNIE allows us to evaluate the mediator-outcome relationship at both the reference and exposed levels of temperature, thereby capturing potential interaction between ozone (mediator) and temperature (exposure).

Fourth, in contrast to previous studies that relied on bootstrap confidence intervals, we used samples from the Bayesian posterior predictive distributions to obtain uncertainty intervals. Our application involves time-series data of temperature, ozone and ED visits, violating the independence assumption required for standard bootstrap methods \citep{RN30}. Our Bayesian bootstrap overcomes this by avoiding resampling of temporally correlated observations, while also eliminating the need to repeatedly fit BART on bootstrap samples, which is highly computationally demanding. 

In future work, we plan to expand our method in the following ways. First, we assumed a log-linear relationship between ozone and number of ED visits, although the true association may be more complex. Allowing for nonlinearity in both the outcome and mediator regression could further improve the robustness of our method to strict parametric assumptions. Second, we did not incorporate spatial information into our procedure. Spatial heterogeneity of ozone's mediation effects has previously been observed in a study examining the mediation role of the relationship between heat waves and mortality conducted in France \citep{RN18}. To address this, we could apply our method to multiple regions separately, or incorporate spatial heterogeneity into BART through spatial covariates \citep{RN36}. Third, we only included a single mediator in our analysis. A future direction is to evaluate the mediation effects of multiple pollutants (e.g., PM$_{2.5}$) on health outcomes. Previous studies have extended causal mediation frameworks to accommodate multiple mediators, such as sequential mediators \citep{RN37} and longitudinal mediators \citep{RN38}. Building on these developments, our approach could be extended to estimate the causal effects of multiple or interacting pollutants. 
Summarize contributions, limitations, and future work.

\section*{Acknowledgments}

We thank our health data source, the California Office of Statewide Planning and Development, now California Department of Health Care Access and Information, and its contributing hospitals. The contents of this publication including data analysis, interpretation, conclusions derived, and the views expressed herein are solely those of the authors and do not represent the conclusions or official views of the data source listed above. Authorization to release this information does not imply endorsement of this study or its findings by this data source. The data source, their employees, officers, and agents make no representation, warranty, or guarantee as to the accuracy, completeness, currency, or suitability of the information provided here 
\newpage

\bibliographystyle{apalike}
\bibliography{references}

@article{RN14,
   author = {Ahn, Seoyeong and Kang, Cinoo and Oh, Jieun and Yun, Hyewon and Ahn, Sojin and Kim, Ayoung and Kwon, Dohoon and Park, Jinah and Jang, Hyemin and Kim, Ejin},
   title = {Heterogeneous associations between short-term ambient ozone exposure and morbidities from infants to seniors: A nationwide case-crossover study in South Korea},
   journal = {Journal of Hazardous Materials Advances},
   volume = {17},
   pages = {100531},
   ISSN = {2772-4166},
   year = {2025},
   type = {Journal Article}
}

@article{RN18,
   author = {Alari, Anna and Chen, Chen and Schwarz, Lara and Hdansen, Kristen and Chaix, Basile and Benmarhnia, Tarik},
   title = {The role of ozone as a mediator of the relationship between heat waves and mortality in 15 French urban areas},
   journal = {American Journal of Epidemiology},
   volume = {192},
   number = {6},
   pages = {949-962},
   ISSN = {0002-9262},
   year = {2023},
   type = {Journal Article}
}

@article{RN19,
   author = {Bae, Sanghyuk and Lim, Youn-Hee and Oh, Jongmin and Kwon, Ho-Jang},
   title = {Mediation of daily ambient ozone concentration on association between daily mean temperature and mortality in 7 metropolitan cities of Korea},
   journal = {Environment international},
   volume = {178},
   pages = {108078},
   ISSN = {0160-4120},
   year = {2023},
   type = {Journal Article}
}

@article{RN17,
   author = {Bloomer, Bryan J and Stehr, Jeffrey W and Piety, Charles A and Salawitch, Ross J and Dickerson, Russell R},
   title = {Observed relationships of ozone air pollution with temperature and emissions},
   journal = {Geophysical research letters},
   volume = {36},
   number = {9},
   ISSN = {0094-8276},
   year = {2009},
   type = {Journal Article}
}

@article{RN16,
   author = {Doherty, Ruth M and Heal, Mathew R and O’Connor, Fiona M},
   title = {Climate change impacts on human health over Europe through its effect on air quality},
   journal = {Environmental Health},
   volume = {16},
   pages = {33-44},
   year = {2017},
   type = {Journal Article}
}

@article{RN12,
   author = {Filippidou, EC and Koukouliata, AJPHS},
   title = {Ozone effects on the respiratory system},
   journal = {Progress in Health Sciences},
   volume = {1},
   number = {2},
   pages = {144-155},
   ISSN = {2083-1617},
   year = {2011},
   type = {Journal Article}
}

@book{RN15,
   author = {Fowler, David and Amann, Markus and Anderson, Ross and Ashmore, Mike and Cox, Peter and Depledge, Michael and Derwent, Dick and Grennfelt, Peringe and Hewitt, Nick and Hov, Oystein},
   title = {Ground-level ozone in the 21st century: future trends, impacts and policy implications},
   publisher = {The Royal Society},
   ISBN = {0854037136},
   year = {2008},
   type = {Book}
}

@article{RN20,
   author = {Gong, Xing and Sun, Fengxia and Wei, Li and Zhang, Yi and Xia, Minjie and Ge, Ming and Xiong, Lilin},
   title = {Association of Ozone and Temperature with Ischemic Heart Disease Mortality Risk: Mediation and Interaction Analyses},
   journal = {Environmental Science \& Technology},
   volume = {58},
   number = {46},
   pages = {20378-20388},
   ISSN = {0013-936X},
   year = {2024},
   type = {Journal Article}
}

@article{RN3,
   author = {Hajat, Shakoor and Kosatky, Tom},
   title = {Heat-related mortality: a review and exploration of heterogeneity},
   journal = {Journal of Epidemiology \& Community Health},
   volume = {64},
   number = {9},
   pages = {753-760},
   ISSN = {0143-005X},
   year = {2010},
   type = {Journal Article}
}

@article{RN27,
   author = {Hill, Jennifer and Linero, Antonio and Murray, Jared},
   title = {Bayesian additive regression trees: A review and look forward},
   journal = {Annual Review of Statistics and Its Application},
   volume = {7},
   number = {1},
   pages = {251-278},
   ISSN = {2326-8298},
   year = {2020},
   type = {Journal Article}
}

@article{RN28,
   author = {Kahle, Juliette J and Neas, Lucas M and Devlin, Robert B and Case, Martin W and Schmitt, Michael T and Madden, Michael C and Diaz-Sanchez, David},
   title = {Interaction effects of temperature and ozone on lung function and markers of systemic inflammation, coagulation, and fibrinolysis: a crossover study of healthy young volunteers},
   journal = {Environmental health perspectives},
   volume = {123},
   number = {4},
   pages = {310-316},
   ISSN = {0091-6765},
   year = {2015},
   type = {Journal Article}
}

@article{RN7,
   author = {Knowlton, Kim and Rotkin-Ellman, Miriam and King, Galatea and Margolis, Helene G and Smith, Daniel and Solomon, Gina and Trent, Roger and English, Paul},
   title = {The 2006 California heat wave: impacts on hospitalizations and emergency department visits},
   journal = {Environmental health perspectives},
   volume = {117},
   number = {1},
   pages = {61-67},
   ISSN = {0091-6765},
   year = {2009},
   type = {Journal Article}
}

@article{RN30,
   author = {Künsch, Hans R},
   title = {The jackknife and the bootstrap for general stationary observations},
   journal = {The annals of Statistics},
   pages = {1217-1241},
   ISSN = {0090-5364},
   year = {1989},
   type = {Journal Article}
}

@article{RN4,
   author = {Lay, CR and Mills, Dave and Belova, Anna and Sarofim, MC and Kinney, PL and Vaidyanathan, Ambarish and Jones, Russel and Hall, Robert and Saha, Shubhayu},
   title = {Emergency department visits and ambient temperature: Evaluating the connection and projecting future outcomes},
   journal = {GeoHealth},
   volume = {2},
   number = {6},
   pages = {182-194},
   ISSN = {2471-1403},
   year = {2018},
   type = {Journal Article}
}

@article{RN9,
   author = {Lee, Hyewon and Yoon, Hee-Young},
   title = {Impact of ambient temperature on respiratory disease: a case-crossover study in Seoul},
   journal = {Respiratory Research},
   volume = {25},
   number = {1},
   pages = {73},
   ISSN = {1465-993X},
   year = {2024},
   type = {Journal Article}
}

@article{RN11,
   author = {Lin, Shao and Hsu, Wan-Hsiang and Van Zutphen, Alissa R and Saha, Shubhayu and Luber, George and Hwang, Syni-An},
   title = {Excessive heat and respiratory hospitalizations in New York State: estimating current and future public health burden related to climate change},
   journal = {Environmental health perspectives},
   volume = {120},
   number = {11},
   pages = {1571-1577},
   ISSN = {0091-6765},
   year = {2012},
   type = {Journal Article}
}

@article{RN8,
   author = {Lin, Shao and Luo, Ming and Walker, Randi J and Liu, Xiu and Hwang, Syni-An and Chinery, Robert},
   title = {Extreme high temperatures and hospital admissions for respiratory and cardiovascular diseases},
   journal = {Epidemiology},
   volume = {20},
   number = {5},
   pages = {738-746},
   ISSN = {1044-3983},
   year = {2009},
   type = {Journal Article}
}

@article{RN2,
   author = {Lin, Szu Yu and Ng, Chris Fook Sheng and Kim, Yoonhee and Htay, Zin Wai and Cao, Alton Quan and Pan, Rui and Hashizume, Masahiro},
   title = {Ambient temperature and nervous system diseases-related mortality in Japan from 2010 to 2019: a time-stratified case-crossover analysis},
   journal = {Science of The Total Environment},
   volume = {867},
   pages = {161464},
   ISSN = {0048-9697},
   year = {2023},
   type = {Journal Article}
}

@article{RN10,
   author = {Ma, Yuxia and Zhou, Jianding and Yang, Sixu and Yu, Zhiang and Wang, Fei and Zhou, Ji},
   title = {Effects of extreme temperatures on hospital emergency room visits for respiratory diseases in Beijing, China},
   journal = {Environmental Science and Pollution Research},
   volume = {26},
   pages = {3055-3064},
   ISSN = {0944-1344},
   year = {2019},
   type = {Journal Article}
}

@article{RN13,
   author = {Magzamen, Sheryl and Moore, Brianna F and Yost, Michael G and Fenske, Richard A and Karr, Catherine J},
   title = {Ozone-related respiratory morbidity in a low-pollution region},
   journal = {Journal of occupational and environmental medicine},
   volume = {59},
   number = {7},
   pages = {624-630},
   ISSN = {1076-2752},
   year = {2017},
   type = {Journal Article}
}

@article{RN1,
   author = {Moghadamnia, Mohammad Taghi and Ardalan, Ali and Mesdaghinia, Alireza and Keshtkar, Abbas and Naddafi, Kazem and Yekaninejad, Mir Saeed},
   title = {Ambient temperature and cardiovascular mortality: a systematic review and meta-analysis},
   journal = {PeerJ},
   volume = {5},
   pages = {e3574},
   ISSN = {2167-8359},
   year = {2017},
   type = {Journal Article}
}

@article{RN23,
   author = {Monaghan, Andrew J and Hu, Leiqiu and Brunsell, Nathaniel A and Barlage, Michael and Wilhelmi, Olga V},
   title = {Evaluating the impact of urban morphology configurations on the accuracy of urban canopy model temperature simulations with MODIS},
   journal = {Journal of Geophysical Research: Atmospheres},
   volume = {119},
   number = {11},
   pages = {6376-6392},
   ISSN = {2169-897X},
   year = {2014},
   type = {Journal Article}
}

@article{RN21,
  title = {The {{High-resolution Urban Meteorology}} for {{Impacts Dataset}} ({{HUMID}}) Daily for the {{Conterminous United States}}},
  author = {Newman, Andrew J. and Kalb, Christina and Chakraborty, T. C. and Fitch, Amy and Darrow, Lyndsey A. and Warren, Joshua L. and Strickland, Matthew J. and Holmes, Heather A. and Monaghan, Andrew J. and Chang, Howard H.},
  year = 2024,
  month = dec,
  journal = {Scientific Data},
  volume = {11},
  number = {1},
  pages = {1321},
  publisher = {Nature Publishing Group},
  issn = {2052-4463},
  doi = {10.1038/s41597-024-04086-2},
  urldate = {2026-02-18},
  copyright = {2024 The Author(s)},
  langid = {english}
}

@inbook{RN25,
    author = {Pearl, Judea},
   title = {Direct and indirect effects},
   booktitle = {Probabilistic and causal inference: the works of Judea Pearl},
   pages = {373-392},
   year = {2022},
   type = {Book Section},
  month = jan,
  edition = {1},
  volume = {36},
  publisher = {Association for Computing Machinery},
  address = {New York, NY, USA},
  isbn = {978-1-4503-9586-1}
}

@article{RN24,
   author = {Senthilkumar, Niru and Gilfether, Mark and Metcalf, Francesca and Russell, Armistead G and Mulholland, James A and Chang, Howard H},
   title = {Application of a fusion method for gas and particle air pollutants between observational data and chemical transport model simulations over the contiguous United States for 2005–2014},
   journal = {International journal of environmental research and public health},
   volume = {16},
   number = {18},
   pages = {3314},
   ISSN = {1660-4601},
   year = {2019},
   type = {Journal Article}
}

@article{RN6,
   author = {Strand, Linn B and Barnett, Adrian G and Tong, Shilu},
   title = {The influence of season and ambient temperature on birth outcomes: a review of the epidemiological literature},
   journal = {Environmental research},
   volume = {111},
   number = {3},
   pages = {451-462},
   ISSN = {0013-9351},
   year = {2011},
   type = {Journal Article}
}

@article{RN29,
   author = {VanderWeele, Tyler J and Vansteelandt, Stijn},
   title = {Odds ratios for mediation analysis for a dichotomous outcome},
   journal = {American journal of epidemiology},
   volume = {172},
   number = {12},
   pages = {1339-1348},
   ISSN = {1476-6256},
   year = {2010},
   type = {Journal Article}
}

@article{RN26,
   author = {Varotsos, Konstantinos V and Giannakopoulos, Christos and Tombrou, Maria},
   title = {Ozone-temperature relationship during the 2003 and 2014 heatwaves in Europe},
   journal = {Regional Environmental Change},
   volume = {19},
   pages = {1653-1665},
   ISSN = {1436-3798},
   year = {2019},
   type = {Journal Article}
}

@article{RN5,
   author = {Winquist, Andrea and Grundstein, Andrew and Chang, Howard H and Hess, Jeremy and Sarnat, Stefanie Ebelt},
   title = {Warm season temperatures and emergency department visits in Atlanta, Georgia},
   journal = {Environmental research},
   volume = {147},
   pages = {314-323},
   ISSN = {0013-9351},
   year = {2016},
   type = {Journal Article}
}

@article{RN31,
  title={Bronchoconstriction triggered by breathing hot humid air in patients with asthma: role of cholinergic reflex},
  author={Hayes Jr, Don and Collins, Paul B and Khosravi, Mehdi and Lin, Ruei-Lung and Lee, Lu-Yuan},
  journal={American journal of respiratory and critical care medicine},
  volume={185},
  number={11},
  pages={1190--1196},
  year={2012},
  publisher={American Thoracic Society}
}

@article{RN32,
  title={Heat stress in older individuals and patients with common chronic diseases},
  author={Kenny, Glen P and Yardley, Jane and Brown, Candice and Sigal, Ronald J and Jay, Ollie},
  journal={Cmaj},
  volume={182},
  number={10},
  pages={1053--1060},
  year={2010},
  publisher={CMAJ}
}

@article{RN33,
  title={Physiological factors characterizing heat-vulnerable older adults: a narrative review},
  author={Meade, Robert D and Akerman, Ashley P and Notley, Sean R and McGinn, Ryan and Poirier, Paul and Gosselin, Pierre and Kenny, Glen P},
  journal={Environment international},
  volume={144},
  pages={105909},
  year={2020},
  publisher={Elsevier}
}

@article{RN34,
  title={Time series analysis on the health effects of temperature: advancements and limitations},
  author={Gasparrini, Antonio and Armstrong, Ben},
  journal={Environmental research},
  volume={110},
  number={6},
  pages={633--638},
  year={2010},
  publisher={Elsevier}
}

@article{RN36,
  title={BART-SIMP: a novel framework for flexible spatial covariate modeling and prediction using Bayesian additive regression trees},
  author={Jiang, Alex Ziyu and Wakefield, Jon},
  journal={arXiv preprint arXiv:2309.13270},
  year={2023}
}

@article{RN37,
  title={Flexible mediation analysis with multiple mediators},
  author={Steen, Johan and Loeys, Tom and Moerkerke, Beatrijs and Vansteelandt, Stijn},
  journal={American journal of epidemiology},
  volume={186},
  number={2},
  pages={184--193},
  year={2017},
  publisher={Oxford University Press}
}

@article{RN38,
  title={Mediation analysis of time-to-event endpoints accounting for repeatedly measured mediators subject to time-varying confounding},
  author={Vansteelandt, Stijn and Linder, Martin and Vandenberghe, Sjouke and Steen, Johan and Madsen, Jesper},
  journal={Statistics in medicine},
  volume={38},
  number={24},
  pages={4828--4840},
  year={2019},
  publisher={Wiley Online Library}
}

@article{RN39,
  title={The synergistic and mediating effects of ozone on associations between high temperature, heatwaves and mortality in the Greater London area between 2010 and 2018},
  author={Gao, Juan and Wood, Dylan and Katsouyanni, Klea and Benmarhnia, Tarik and Evangelopoulos, Dimitris},
  journal={Environmental Research},
  volume={277},
  pages={121577},
  year={2025},
  publisher={Elsevier}
}

@article{RN40,
  title={Ozone serve as mediator and effect modifier in the temperature-eGFR association: A longitudinal study on health examination cohort},
  author={Huang, Zhongguo and Lu, Jinlu and He, Guanhao and Hu, Jianxiong and Guo, Xuming and Chen, Mingxuan and Liu, Tao and Lin, Shao and Liu, Fanna and Xu, Ying and others},
  journal={Environmental Research},
  pages={122578},
  year={2025},
  publisher={Elsevier}
}

@article{andersonHeatrelatedEmergencyHospitalizations2013,
  title = {Heat-Related {{Emergency Hospitalizations}} for {{Respiratory Diseases}} in the {{Medicare Population}}},
  author = {Anderson, G. Brooke and Dominici, Francesca and Wang, Yun and McCormack, Meredith C. and Bell, Michelle L. and Peng, Roger D.},
  year = 2013,
  month = may,
  journal = {American Journal of Respiratory and Critical Care Medicine},
  volume = {187},
  number = {10},
  pages = {1098--1103},
  publisher = {American Thoracic Society - AJRCCM},
  issn = {1073-449X},
  doi = {10.1164/rccm.201211-1969OC},
  urldate = {2026-02-18}
}

@article{zhuHeatExposureRespiratory2025,
  title = {Heat Exposure and Respiratory Diseases Health Outcomes: {{An}} Umbrella Review},
  shorttitle = {Heat Exposure and Respiratory Diseases Health Outcomes},
  author = {Zhu, Zhenggang and Ji, Binbin and Tian, Jun and Yin, Ping},
  year = 2025,
  month = mar,
  journal = {Science of The Total Environment},
  volume = {970},
  pages = {179052},
  issn = {0048-9697},
  doi = {10.1016/j.scitotenv.2025.179052},
  urldate = {2026-02-18}
}

@article{lineroMediationAnalysisUsing2025,
  title = {Mediation Analysis Using {{Bayesian}} Tree Ensembles.},
  author = {Linero, Antonio R. and Zhang, Qian},
  year = 2025,
  month = feb,
  journal = {Psychological Methods},
  volume = {30},
  number = {1},
  pages = {60--82},
  publisher = {American Psychological Association},
  issn = {1939-1463},
  doi = {10.1037/met0000504},
  langid = {english}
}

@article{tingEstimatingHeterogeneousCausal2025,
  title = {Estimating {{Heterogeneous Causal Mediation Effects}} with {{Bayesian Decision Tree Ensembles}}},
  author = {Ting, Angela and Linero, Antonio R.},
  year = 2025,
  month = jul,
  journal = {Journal of the American Statistical Association},
  volume = {120},
  number = {551},
  pages = {1400--1413},
  publisher = {Taylor \& Francis},
  issn = {0162-1459},
  doi = {10.1080/01621459.2025.2491155}
}

\clearpage

\appendix
\renewcommand{\thetable}{S\arabic{table}}
\renewcommand{\thefigure}{S\arabic{figure}}
\setcounter{table}{0}
\setcounter{figure}{0}
\setcounter{page}{1}
\pagenumbering{arabic}

\part*{Appendix}
\addcontentsline{toc}{part}{Appendix}

\setcounter{lemma}{0}
\setcounter{theorem}{0}
\setcounter{example}{0}
\setcounter{prop}{0}
\setcounter{corollary}{0}
\setcounter{definition}{0}
\setcounter{remark}{0}

\renewcommand{\thelemma}{S\arabic{lemma}}
\renewcommand{\thetheorem}{S\arabic{theorem}}
\renewcommand{\theexample}{S\arabic{example}}
\renewcommand{\theprop}{S\arabic{prop}}
\renewcommand{\thecorollary}{S\arabic{corollary}}
\renewcommand{\thedefinition}{S\arabic{definition}}
\renewcommand{\theremark}{S\arabic{remark}}

\etocsettocstyle{}{}
{\small
\localtableofcontents
}

\section{Estimator of Causal Mediation Effects}\label{sec:appendix_estimation}

The marginal expectation of the nested counterfactual variable can be written as the integral of conditional expectation over $\bm{c}$:
    \begin{align}
    \text{E}[Y(x,M(x^*))] = \int \text{E}[Y(x,M(x^*)) \mid \boldsymbol{C}] f(\bm{c})\, d\bm{c}
    \end{align}
Once the conditional expectation is estimated, the marginal expectation can be obtained by Monte Carlo approximation.

\subsection{Spline approach}

When an additive spline regression model is used to capture the associations between temperature and ozone concentration:
    \begin{align*}
    \text{E}( M | X, \bm{C}) = \beta_0 + g(x,\bm{\beta_1}) + \bm{\beta_2^\top c}\,.
    \end{align*}
By the mediator regression and outcome regression, the expectation of the nested counterfactual variable given $\bm{C}$ can be written as:
    \begin{align*}
    \mathbb{E}[Y(x,M(x^*)) \mid \boldsymbol{C}] &= \exp{(\theta_0 + f(x, \boldsymbol{\theta_1}) +\boldsymbol{\theta_{4}^\top c})} \ \mathbb{E}[\exp{(\theta_2M+\theta_{3x}M)} \mid X=x^* ,\boldsymbol{C}] \\
    & = \exp(\theta_0 + f(x, \boldsymbol{\theta_1}) +\boldsymbol{\theta_{4}^\top c})  \\ 
    & \qquad \times  \exp[(\theta_2 + \theta_{3x})(\beta_0 + g(x^*, \boldsymbol{\beta_1})+ \boldsymbol{\beta_2^\top c})+\sigma^2 (\theta_2 + \theta_{3x})^2/2].
    \end{align*}
Where $\sigma$ is the residual error for $M$ in the linear regression model. Then we can get the four causal mediation effects in the multiplicative scale. The pure natural direct effect is as follows:
    \begin{align*}
    \text{PNDE}^{(t)} &= \frac{\text{E}[Y(x,M(x^*)) \mid \boldsymbol{c}_{t}]}{\text{E}[Y(x^*,M(x^*)) \mid \boldsymbol{c}_{t}]} \\
    & = \frac{\exp{(\theta_0 + f(x, \boldsymbol{\theta_1})+\boldsymbol{\theta_{4}^\top c_{t}})} \ \exp{[(\theta_2 + \theta_{3x})(\beta_0 + g(x^*, \boldsymbol{\beta_1})+ \boldsymbol{\beta_2^\top c_{t}})+\sigma^2 (\theta_2 + \theta_{3x})^2/2]}}{\exp{(\theta_0 + f(x^*, \boldsymbol{\theta_1})+\boldsymbol{\theta_4c_{t}})} \ \exp{[(\theta_2 + \theta_{3x^*})(\beta_0 + g(x^*, \boldsymbol{\beta_1}) + \boldsymbol{\beta_2^\top c_{t}})+\sigma^2 (\theta_2 + \theta_{3x^*})^2/2]}} \\
    & = \exp{[(f(x, \boldsymbol{\theta_1}) - f(x^*, \boldsymbol{\theta_1}))]} \\
    & \ \ \ \times \exp{[(\theta_{3x} - \theta_{3x^*})(\beta_0 + g(x^*, \boldsymbol{\beta_1}) + \boldsymbol{\beta_2^\top c_t}) + \sigma^2 \theta_2 (\theta_{3x} - \theta_{3x^*}) + \sigma^2 (\theta_{3x}^2 - \theta_{3x^*}^2)/2]}, \\
    \text{PNDE} & = \sum_{t=1}^{N} \text{PNDE}^{(t)},
    \end{align*}
which is the Monte Carlo approximation for PNDE (approximation for integral over $\bm{c_t}$). The total natural direct effect is as follows:
    \begin{align*}
    \text{TNDE}^{(t)} &= \frac{\text{E}[Y(x,M(x)) \mid \boldsymbol{c}_{t}]}{\text{E}[Y(x^*,M(x)) \mid \boldsymbol{c}_{t}]} \\
    & = \frac{\exp{(\theta_0 + f(x, \boldsymbol{\theta_1})+\boldsymbol{\theta_{4}^\top c_{t}})} \ \exp{[(\theta_2 + \theta_{3x})(\beta_0 + g(x, \boldsymbol{\beta_1})+ \boldsymbol{\beta_2^\top c_{t}})+\sigma^2 (\theta_2 + \theta_{3x})^2/2]}}{\exp{(\theta_0 + f(x^*, \boldsymbol{\theta_1})+\boldsymbol{\theta_4c_{t}})} \ \exp{[(\theta_2 + \theta_{3x^*})(\beta_0 + g(x, \boldsymbol{\beta_1}) + \boldsymbol{\beta_2^\top c_{t}})+\sigma^2 (\theta_2 + \theta_{3x^*})^2/2]}} \\
    & = \exp{[(f(x, \boldsymbol{\theta_1}) - f(x^*, \boldsymbol{\theta_1}))]} \\
    & \ \ \ \times \exp{[(\theta_{3x} - \theta_{3x^*})(\beta_0 + g(x, \boldsymbol{\beta_1}) + \boldsymbol{\beta_2^\top c_t}) + \sigma^2 \theta_2 (\theta_{3x} - \theta_{3x^*}) + \sigma^2 (\theta_{3x}^2 - \theta_{3x^*}^2)/2]}, \\
    \text{TNDE} & = \sum_{t=1}^{N} \text{TNDE}^{(t)},
    \end{align*}
which is the Monte Carlo approximation for TNDE. The total natural direct effect is as follows:
    \begin{align*}
    \text{TNIE} &= \frac{\text{E}[Y(x,M(x)) \mid \boldsymbol{c_t}]}{\text{E}[Y(x,M(x^*)) \mid \boldsymbol{c_t}]} \\
    & = \frac{\exp{(\theta_0 + f(x, \boldsymbol{\theta_1}) +\boldsymbol{\theta_4 c_{t}})} \ \exp{[(\theta_2 + \theta_{3x})(\beta_0 + g(x, \boldsymbol{\beta_1}) + \boldsymbol{\beta_2^\top c_{t}})+\sigma^2 (\theta_2 + \theta_{3x})^2/2]}}{\exp{(\theta_0 + f(x, \boldsymbol{\theta_1})+\boldsymbol{\theta_4 c_{t}})} \ \exp{[(\theta_2 + \theta_{3x})(\beta_0 + g(x^*, \boldsymbol{\beta_1}) + \boldsymbol{\beta_2^\top c_{t}})+\sigma^2 (\theta_2 + \theta_{3x})^2/2]}} \\
    & = \exp{[(\theta_2 + \theta_{3x})(g(x, \boldsymbol{\beta_1}) - g(x^*, \boldsymbol{\beta_1}))]}.
    \end{align*}
As $\bm{c}$ is canceled out, TNIE can be obtained directly without using Monte Carlo approximation. Similarly, the pure natural direct effect is as follows:
    \begin{align*}
    \text{PNIE} &= \frac{\text{E}[Y(x^*,M(x)) \mid \boldsymbol{c_t}]}{\text{E}[Y(x^*,M(x^*)) \mid \boldsymbol{c_t}]} \\
    & = \frac{\exp{(\theta_0 + f(x^*, \boldsymbol{\theta_1}) +\boldsymbol{\theta_4 c_{t}})} \ \exp{[(\theta_2 + \theta_{3x^*})(\beta_0 + g(x, \boldsymbol{\beta_1}) + \boldsymbol{\beta_2^\top c_{t}})+\sigma^2 (\theta_2 + \theta_{3x^*})^2/2]}}{\exp{(\theta_0 + f(x^*, \boldsymbol{\theta_1})+\boldsymbol{\theta_4 c_{t}})} \ \exp{[(\theta_2 + \theta_{3x^*})(\beta_0 + g(x^*, \boldsymbol{\beta_1}) + \boldsymbol{\beta_2^\top c_{t}})+\sigma^2 (\theta_2 + \theta_{3x^*})^2/2]}} \\
    & = \exp{[(\theta_2 + \theta_{3x^*})(g(x, \boldsymbol{\beta_1}) - g(x^*, \boldsymbol{\beta_1}))]}.
    \end{align*}
The confidence interval can be obtained by parametric bootstrapping.

\subsection{BART-based approach}
When a BART is used to capture the associations between temperature and ozone concentration, we have:
    \begin{align*}
    \text{E}[Y(X,M({x}^*)) \mid \boldsymbol{C}] 
    & = \exp{(\theta_0 + f(x, \boldsymbol{\theta_1}) +\boldsymbol{\theta_{4}^\top c})} \times \text{E}[\exp{(\theta_2m+\theta_{3x}m)} \mid X=x^* ,\boldsymbol{C}] \\
    \end{align*}
Let:
$$\text{E}[Y(\boldsymbol{x},M(\boldsymbol{x}^*)) \mid \boldsymbol{C}] = \mathcal{F} (x, x^* \mid \boldsymbol{C}).$$
To avoid fitting BART models for thousands of times when employing a parametric bootstrap, we used samples from Bayesian posterior prediction to get the estimation and the confidence interval:
    \begin{enumerate}
    \item Generate $K$ set of coefficients of the health model by $\boldsymbol{\hat{\theta}}^{(k)} \sim \mathcal{MVN}(\boldsymbol{\hat{\theta}}, \Sigma)$ from the asymptotic distribution of the coefficients, where $K = 5000$.
    \item Fit the BART model with real data. Generate $K = 5000$ sets of predicted mediators when the exposure is at reference level ($X=x^*$) for every observation by the fitted BART: $$\hat{m}_{|x^*}^{(k,t)} \sim \mathcal{N}(\sum_{g=1}^G \mathcal{T}_g (X = x^*, \bm{C}_t)^{(k)}, \sigma^2).$$ The estimated standard deviance $\hat{\sigma}^{(k)}$ can be obtained from the BART model for each observation as well. 
    \item In the $k^{th}$ iteration:
        \begin{itemize}
    	\item Plug in $\boldsymbol{\hat{\theta}}^{(k)}$ and $\hat{m}_{x^*}^{(k,t)}$ into $\mathcal{F} (x, x^* \mid \boldsymbol{C})$ for every observation:
            \begin{align*}
            &\hat{\mathcal{F}}(x,x^*)^{(k,t)} = \hat{\mathcal{F}}(x,x^* \mid \boldsymbol{c}_{t})^{(k)} \\
            &\  = \exp{(\theta_0 + f(x,\boldsymbol{\hat{\theta}_1^{(k)}}) +\boldsymbol{\hat{\theta}_4^{(k) \top}c_t})}\exp\left( (\hat{\theta}_2^{(k)} + \hat{\theta}_{3x}^{(k)}) \hat{m}_{|x^*}^{(k,t)} + \frac{[\hat{\sigma}^{(k)}(\hat{\theta}_2^{(k)} + \hat{\theta}_{3x}^{(k)})]^2}{2} \right).
            \end{align*}
        The estimations for the $k^{th}$ iteration are:
            \begin{align*}
            &\boldsymbol{\hat{\mathcal{F}}}(x,x^*)^{(k)} = (\hat{\mathcal{F}}(x,x^* \mid \boldsymbol{c}_1)^{(k)}, \hat{\mathcal{F}}(x,x^* \mid \boldsymbol{c}_2)^{(k)}...\hat{\mathcal{F}}(x,x^* \mid \boldsymbol{c}_T)^{(k)})
            \end{align*}
        \item Generate weights $\boldsymbol{w} = \{w_1, w_2...w_t...w_T\}'$ for every observation by a Dirichlet Distribution $\boldsymbol{w} \sim \mathcal{DIR} (\alpha = (1, 1, ..,1)_{(1 \times T)})$.
        \item By assigning weight to every observation, we are doing a Bayesian Bootstrap to resample the data and get the average of the mediation formula estimation over $t$:
            \begin{align*}
            &\hat{\mathcal{F}}(x,x^*)^{(k)}_{avg} = \boldsymbol{w^\top}  \boldsymbol{\hat{\mathcal{F}}}(x,x^*)^{(k)} 
            \end{align*}
        \end{itemize}
     \item By repeating step 3 for $K$ times, we could get the posterior sample for the mediation formula, including $\hat{\mathcal{F}}(x,x^*)^{(k)}_{avg}$, $\hat{\mathcal{F}}(x,x)^{(k)}_{avg}$, $\hat{\mathcal{F}}(x^*,x^*)^{(k)}_{avg}$, and $\hat{\mathcal{F}}(x^*,x)^{(k)}_{avg}$. The confidence interval and point estimate can be obtained from the posterior draw. Take PNDE as an example:
    $$\textbf{PNDE} = [\frac{\hat{\mathcal{F}}(x,x^*)^{(1)}_{avg}}{\hat{\mathcal{F}}(x^*,x^*)^{(1)}_{avg}},\frac{\hat{\mathcal{F}}(x,x^*)^{(2)}_{avg}}{\hat{\mathcal{F}}(x^*,x^*)^{(2)}_{avg}}...\frac{\hat{\mathcal{F}}(x,x^*)^{(K)}_{avg}}{\hat{\mathcal{F}}(x^*,x^*)^{(K)}_{avg}}],$$
    the point estimate is mean of $\textbf{PNDE}$ and the $0.025$ and $0.975$ quantile of $\textbf{PNDE}$ are the lower bound and the upper bound of the confidence interval.
    \end{enumerate}
    
\section{Simulation Data Generating Process}\label{sec:appendix_sim}
\subsection{Fixed covariate design and exposure contrasts}

We calibrated our simulation to the observed Los Angeles data. Our intention was to keep all parts of the data generating process as compatible with the dataset as possible. The design comprised $n=2{,}208$ daily observations from May through October in 2005 to 2016. For day $i$, let $X_i$ denote maximum temperature and let $C_i=(H_i,D_i,R_i,W_i,J_i)$ contain specific humidity, day of year, calendar year, weekday, and the holiday indicator. The 2,208 observed rows $(X_i,C_i)$ were held fixed in every Monte Carlo replicate.

The reference exposure $x^*$ was the observed median temperature, and the comparison exposures were its observed 75th, 85th, and 95th percentiles. These values were:

\begin{center}
\begin{tabular}{lrrrrrr}
\toprule
Quantile        & 0.50 & 0.65 & 0.75 & 0.80 & 0.85 & 0.95 \\
\midrule
Temperature (K) & 302.311 & 303.831 & 304.921 & 305.694 & 306.466 & 308.766 \\
\bottomrule
\end{tabular}
\end{center}
The 0.65 and 0.80 quantiles were used only to define the nonadditive mediator scenario below. Temperature quartiles were defined once from the observed temperature distribution, and the same cut points were used for factual and counterfactual temperatures.

\subsection{Ozone mediator data-generating models}

Let $B_X(x)$, $B_H(h)$, and $B_D(d)$ be natural cubic spline bases with six degrees of freedom. Their knots and boundary knots were determined from the observed temperature, humidity, and day-of-year distributions, respectively, and were held fixed. We formed the design vector
\begin{equation*}
Z_M(x,c)
=
\left[
1,\ B_X(x),\ B_H(h),\ j,\ w,\ B_D(d),\ r,\ 
B_D(d)\mathbin{\otimes}r
\right],
\end{equation*}
where weekday and year were represented as factors and $B_D(d)\mathbin{\otimes}r$ denotes the day-of-year spline by year interactions. We fit the linear model
\begin{equation*}
M_i^{\mathrm{obs}}
=
Z_M(X_i,C_i)\beta_M+\varepsilon_i
\end{equation*}
to the observed ozone variable. The fitted coefficients $\widehat\beta_M$ and residual variance

\begin{equation*}
\widehat\sigma_M^2=0.127245
\end{equation*}
were treated as fixed true parameters in the simulation. We considered two mediator data-generating models. In the additive scenario,
\begin{align*}
\mu_{\mathrm{add}}(x,c)
&=
Z_M(x,c)\widehat\beta_M, \\
M_i\mid X_i,C_i
&\sim
N\!\left\{
\mu_{\mathrm{add}}(X_i,C_i),
\widehat\sigma_M^2
\right\}.
\end{align*}
This scenario included nonlinear additive effects of temperature and humidity but no interaction between them. It was therefore correctly specified by the additive natural spline mediator model used as the parametric comparator.

In the nonadditive scenario, we added a bounded temperature-by-humidity threshold interaction. Let $q_a$ denote the observed $a$th temperature quantile, let $\widetilde H$ denote median humidity, and define
\begin{equation}
\begin{aligned}
r_X(x)
&=
0.65{I}(x>q_{0.65})
+
0.35{I}(x>q_{0.80}), \\
r_H(h)
&=
2{I}(h>\widetilde H)-1, \\
\delta(x,h)
&=
\widehat\sigma_M r_X(x)r_H(h).
\end{aligned}
\label{eq:interaction}
\end{equation}
The nonadditive/interaction mediator mean and sampling distribution were
\begin{align*}
\mu_{\mathrm{int}}(x,c)
&=
Z_M(x,c)\widehat\beta_M+\delta(x,h), \\
M_i\mid X_i,C_i
&\sim
N\!\left\{
\mu_{\mathrm{int}}(X_i,C_i),
\widehat\sigma_M^2
\right\}.
\end{align*}

The additive component and residual variance were identical in the two scenarios. The interaction described in Equation~\eqref{eq:interaction} was their only difference. The interaction was zero at the median reference exposure because $q_{0.50}$ was below both thresholds. Above $q_{0.80}$, its absolute magnitude was one mediator residual standard deviation ($0.357$), with opposite signs below and above median humidity. This designed interaction was the only component that was not obtained by fitting the
observed data.

\subsection{ED visit outcome data-generating model}

Let $Q(x)\in\{1,2,3,4\}$ indicate the quartile containing temperature $x$. Conditional on temperature, the mediator, and the fixed covariates, the outcome mean followed the log-linear model

\begin{equation}
\begin{aligned}
\log\{\mu_Y(x,m,c)\}
={}&
Z_M(x,c)^{\mathsf T}\theta_C
+\theta_Mm
+\sum_{k=2}^{4}
\theta_{Mk}m {I}\{Q(x)=k\}.
\end{aligned}
\label{eq:outcome-mean}
\end{equation}
Therefore the outcome model used the same six df natural spline terms and calendar adjustment as the additive mediator model and allowed the mediator coefficient to differ across temperature quartiles. We estimated Equation~\eqref{eq:outcome-mean} once in the observed data using a quasi-Poisson model with a log link. All fitted regression coefficients were then held fixed as the true outcome coefficients. In particular, the fitted mediator main coefficient and its quartile-specific modifiers were

\begin{equation*}
\left(
\widehat\theta_M,
\widehat\theta_{M2},
\widehat\theta_{M3},
\widehat\theta_{M4}
\right)
=
\left(
0.0130491,
-0.0020327,
-0.0012216,
-0.0085876
\right).
\end{equation*}

The estimated quasi-Poisson dispersion was $\widehat\phi=2.81178$. Outcomes were generated from a negative-binomial distribution parameterized to retain the conditional mean and variance:
\begin{align*}
Y_i\mid X_i,M_i,C_i
&\sim
\operatorname{NegBin}\!\left(\mu_{Yi},s_i\right), \\
\mu_{Yi}
&=
\mu_Y(X_i,M_i,C_i), \\
s_i
&=
\frac{\mu_{Yi}}{\widehat\phi-1}.
\end{align*}
Under this parameterization, $\operatorname{Var}(Y_i\mid X_i,M_i,C_i)=\widehat\phi\mu_{Yi}$. Evaluated at the observed mediator and covariate values, the calibrated model had a mean count of $1{,}566.68$, equal to the observed mean up to rounding.

\end{document}